\documentclass{kapproc}

\usepackage{graphicx,amssymb}

\kluwerbib

\normallatexbib

\begin{document}

\articletitle{Magnetic Properties of Cuprate Perovskites}

\author{A. Sherman}
\affil{Institute of Physics, University of Tartu, Riia 142, 51014
Tartu, Estonia}
\email{alexei@fi.tartu.ee}

\begin{abstract}
The magnetic susceptibility of underdoped yttrium and lanthanum
cup\-ra\-tes is interpreted based on the self-consistent solution of
the $t$-$J$ model of a Cu-O plane. The calculations reproduce correctly
the frequency dependencies of the susceptibility in
YBa$_2$Cu$_3$O$_{7-y}$ and La$_{2-x}$Sr$_x$CuO$_4$ attributing their
dissimilarity to the difference in the damping of spin excitations. In
YBa$_2$Cu$_3$O$_{7-y}$ these excitations are well defined at the
antiferromagnetic wave vector ${\bf Q}=(\pi,\pi)$ even in the normal
state which manifests itself in a pronounced maximum -- the resonance
peak -- in the susceptibility. In La$_{2-x}$Sr$_x$CuO$_4$ the spin
excitations are overdamped which leads to a broad
low-fre\-q\-u\-e\-n\-cy feature in the susceptibility. The
low-frequency incommensurability in the magnetic response is attributed
to a dip in the magnon damping at {\bf Q}. The calculated concentration
and temperature dependencies of the incommensurability parameter
conform with experimental observations. Generally the incommensurate
magnetic response is not accompanied with an inhomogeneity of the
carrier density.
\end{abstract}


\section{Introduction}
Among the results obtained with inelastic neutron scattering is the
detailed information on the magnetic susceptibility in
YBa$_2$Cu$_3$O$_{7-y}$ and lanthanum cuprates which reveals a
considerable difference in their magnetic properties. A sharp magnetic
collective mode -- the so-called resonance peak -- was observed at
frequencies $\omega_r=20-40$~meV in YBa$_2$Cu$_3$O$_{7-y}$ and some
other cuprates, in underdoped case both below and above $T_c$
\cite{Bourges}. In the momentum space the mode is strongly peaked at
the antiferromagnetic wave vector ${\bf Q}=(\pi,\pi)$. Contrastingly,
no resonance peak was observed in lanthanum cuprates. Instead for low
temperatures a broad feature was detected at $\omega_f\approx 7$~meV
and for this and lower frequencies the susceptibility was found to be
peaked at incommensurate momenta $(\pi\pm 2\pi\delta,\pi)$ and
$(\pi,\pi\pm 2\pi\delta)$ \cite{Aeppli}. Recently the low-frequency
incommensurate magnetic response was also observed in
YBa$_2$Cu$_3$O$_{7-y}$ \cite{Arai}. Presently the fundamental
difference of the susceptibility frequency dependencies in these two
groups of cuprates is actively debated, as well as the nature of the
resonance peak and the low-frequency magnetic incommensurability.

The aim of this work is to demonstrate that the above-mentioned unusual
properties of cuprates can be interpreted in the framework of the
$t$-$J$ model of a Cu-O plane which is a common structure element of
these crystals. The model was shown to describe correctly the
low-energy part of the spectrum of the realistic extended Hubbard model
\cite{Zhang}. To take proper account of strong electron correlations
inherent in moderately doped cuprate perovskites the description in
terms of Hubbard operators and Mori's projection operator technique
\cite{Mori} are used. The self-energy equations for hole and spin
Green's functions obtained in this approach are self-consistently
solved for the ranges of hole concentrations $0\leq x\lesssim 0.16$ and
temperatures 2~K$\leq T\leq$1200~K. Lattices with 20$\times$20 sites
and larger are used.

The calculations reproduce correctly the frequency and momentum
dependencies of the resonance peak in YBa$_2$Cu$_3$O$_{7-y}$, and its
variation with doping and temperature in normal and superconducting
states. The peak is connected with the excitation branch of localized
Cu spins and the peak frequency is close to the frequency of these
excitations at {\bf Q}. The absence of the resonance peak in lanthanum
cuprates is related to an increased damping of spin excitations which
is possibly connected with a large hole damping in these crystals. For
low frequencies and temperatures the susceptibility is peaked at
incommensurate wave vectors $(\pi\pm 2\pi\delta,\pi)$ and $(\pi,\pi\pm
2\pi\delta)$. The incommensurability is connected with a dip in the
magnon damping at {\bf Q}. In agreement with experiment the
incommensurability parameter $\delta$ is approximately proportional to
$x$ for $0.02\lesssim x\lesssim 0.12$ and saturates for larger
concentrations. The incommensurability disappears with increasing
temperature. Generally the incommensurate magnetic response is not
accompanied by an inhomogeneity of the carrier density.

\section{Main formulas}
The Hamiltonian of the two-dimensional $t$-$J$ model reads
\begin{equation}\label{hamiltonian}
H=\sum_{\bf nm\sigma}t_{\bf nm}a^\dagger_{\bf n\sigma}a_{\bf
m\sigma}+\frac{1}{2}\sum_{\bf nm}J_{\bf nm}\left(s^z_{\bf n}s^z_{\bf
m}+s^{+1}_{\bf n}s^{-1}_{\bf m}\right),
\end{equation}
where $a_{\bf n\sigma}=|{\bf n}\sigma\rangle\langle{\bf n}0|$ is the
hole annihilation operator, {\bf n} and {\bf m} label sites of the
square lattice, $\sigma=\pm 1$ is the spin projection, $|{\bf
n}\sigma\rangle$ and $|{\bf n}0\rangle$ are site states corresponding
to the absence and presence of a hole on the site. These states may be
considered as linear combinations of the products of the $3d_{x^2-y^2}$
copper and $2p_\sigma$ oxygen orbitals of the extended Hubbard model
\cite{Jefferson}. We take into account nearest neighbor interactions
only, $t_{\bf nm}=-t\sum_{\bf a}\delta_{\bf n,m+a}$ and $J_{\bf
nm}=J\sum_{\bf a}\delta_{\bf n,m+a}$ where the four vectors {\bf a}
connect nearest neighbor sites. The spin-$\frac{1}{2}$ operators can be
written as $s^z_{\bf n}=\frac{1}{2}\sum_\sigma\sigma|{\bf
n}\sigma\rangle\langle{\bf n}\sigma|$ and $s^\sigma_{\bf n}=|{\bf
n}\sigma\rangle\langle{\bf n},-\sigma|$.

Properties of the model are determined from the hole and spin retarded
Green's functions
\begin{equation}
G({\bf k}t)=-i\theta(t)\langle\!\{a_{\bf k\sigma}(t),a^\dagger_{\bf
k\sigma}\}\!\rangle, \quad D({\bf k}t)=-i\theta(t)\langle[s^z_{\bf
k}(t),s^z_{\bf -k}]\rangle, \label{green}
\end{equation}
where $a_{\bf k\sigma}$ and
$s^z_{\bf k}$ are the Fourier transforms of the respective site
operators, operator time dependencies and averaging are defined with
the Hamiltonian ${\cal H}=H-\mu\sum_{\bf n}a^\dagger_{\bf
n\sigma}a_{\bf n\sigma}$ with the chemical potential $\mu$. As
mentioned, to obtain self-energy equations for these functions we used
Mori's projection operator technique \cite{Mori,Sherman}. In this
approach the Fourier transform of Green's function $\langle\!\langle
A_0|A^\dagger_0\rangle\!\rangle$ is represented by the continued
fraction
\begin{equation}\label{cfraction}
\langle\!\langle
A_0|A^\dagger_0\rangle\!\rangle=\frac{\displaystyle|A_0\cdot
A_0^\dagger|}{\displaystyle \omega-E_0-\frac{\displaystyle
V_0}{\displaystyle\omega-E_1-\frac{\displaystyle V_1}{\ddots}}}.
\end{equation}
The elements of the fraction $E_i$ and $V_i$ are determined from the
recursive procedure
\begin{eqnarray}
&&[A_n,H]=E_nA_n+A_{n+1}+V_{n-1}A_{n-1},\nonumber\\
&&E_n=|[A_n,H]\cdot A_n^\dagger|\,|A_n\cdot
 A_n^\dagger|^{-1},\label{lanczos}\\
&&V_{n-1}=|A_n\cdot A_n^\dagger|\,|A_{n-1}\cdot
 A_{n-1}^\dagger|^{-1},\quad
 V_{-1}=0,\quad n=0,1,2,\ldots\nonumber
\end{eqnarray}
The operators $A_i$ constructed in the course of this procedure form an
orthogonal set, $|A_i\cdot A^\dagger_j|\propto\delta_{ij}$. In
Eqs.~(\ref{cfraction}) and (\ref{lanczos}) the definition of the inner
product $|A_i\cdot A^\dagger_j|$ depends on the type of the considered
Green's function. For example, for functions (\ref{green}) these are
$\langle\{A_i,A^\dagger_j\}\rangle$ and
$\langle[A_i,A^\dagger_j]\rangle$, respectively. The method described
by Eqs.~(\ref{cfraction}) and (\ref{lanczos}) can be straightforwardly
generalized to the case of many-component operators which is necessary,
for example, to consider Green's functions for Nambu spinors in the
superconducting state \cite{Sherman}.

The residual term of fraction~(\ref{cfraction}) is the Fourier
transform of the quantity
\begin{equation}\label{terminator}
{\cal T}=|A_{nt}\cdot A_n^\dagger|\,|A_{n-1}\cdot
A_{n-1}^\dagger|^{-1},
\end{equation}
where the time evolution of the operator $A_n$ is determined by the
equation
\begin{equation}\label{timevol}
i\frac{d}{dt}{A}_{nt}=\prod_{k=0}^{n-1}(1-P_k)[A_{nt},{\cal H}],\quad
A_{n,t=0}=A_n
\end{equation}
with the projection operators $P_n$ defined as $P_nQ= |Q\cdot
A_n^\dagger|\,|A_n\cdot A_n^\dagger|^{-1}A_n$. The residual term ${\cal
T}$ is a many-particle Green's function which can be estimated by the
decoupling. The decoupling is also used in calculating the elements of
the continued fractions (\ref{cfraction}). Following Ref.~\cite{Kondo}
this approximation is improved by introducing the vertex correction
$\alpha$ which is determined from the constraint of zero site
magnetization
\begin{equation}\label{constraint}
\left\langle s^z_{\bf n}\right\rangle
=\frac{1}{2}\left(1-x\right)-\left\langle s^{-1}_{\bf n}s^{+1}_{\bf
n}\right\rangle=0.
\end{equation}
Since the statistical averaging includes samples with different
ordering orientations, this condition is fulfilled also in magnetically
ordered states.

Equations for Green's functions obtained in this way for the normal
state read \cite{Sherman}
\begin{eqnarray}
G({\bf k}\omega)&=&\frac{\phi}{\omega-\varepsilon_{\bf
 k}+\mu-\Sigma({\bf k}\omega)}, \;
 D({\bf k}\omega)=\frac{4(1-\gamma_{\bf
 k})(J|C_1|+tF_1)}{\omega^2-\omega\Pi({\bf k}\omega)-
 \omega^2_{\bf k}},\nonumber\\
{\rm Im}\,\Sigma({\bf k}\omega)&=&\frac{16\pi t^2}{N\phi}\sum_{\bf
 k'}\int_{-\infty}^\infty d\omega'\biggl[\gamma_{\bf k-k'}+\gamma_{\bf
 k}+{\rm sgn}(\omega')(\gamma_{\bf
 k-k'}-\gamma_{\bf k})\nonumber\\
&\times&\sqrt{\frac{1+\gamma_{\bf k'}}{1-\gamma_{\bf
 k'}}}\biggr]^2 [n_B(-\omega')+n_F(\omega-\omega')]\nonumber\\[-1ex]
&&\label{se}\\[-1ex]
&\times&A({\bf k-k'},\omega-\omega')B({\bf k'}\omega'),\nonumber\\
{\rm Im}\,\Pi({\bf k}\omega)&=&\frac{9\pi t^2J^2(1-x)}{2N(1-\gamma_{\bf
 k})(J|C_1|+tF_1)}\sum_{\bf k'}(\gamma_{\bf k+k'}-\gamma_{\bf
 k'})^2\nonumber\\
&\times&\int^\infty_{-\infty}d\omega' A({\bf k'}\omega')A({\bf k+k'},
 \omega+\omega')\frac{n_F(\omega+\omega')-n_F(\omega')}{\omega},\nonumber
\end{eqnarray}
where $\phi=\frac{1}{2}(1+x)$, $\gamma_{\bf
k}=\frac{1}{2}[\cos(k_x)+\cos(k_y)]$, $C_1=\langle s^{+1}_{\bf
n}s^{-1}_{\bf n+a}\rangle$ and $F_1=\langle a^\dagger_{\bf n}a_{\bf
n+a}\rangle$ are the spin and hole correlations on neighboring sites
which, as well as the hole concentration $x$, can be calculated from
Green's functions,
\begin{equation}\label{energies}
\omega^2_{\bf k}=16J^2\alpha|C_1|(1-\gamma_{\bf k})
(\Delta+1+\gamma_{\bf k}), \; \varepsilon_{\bf k}=-(4\phi
t+6C_1\phi^{-1}t+3F_1\phi^{-1}J)\gamma_{\bf k}
\end{equation}
are the energies of spin excitations and holes with the parameter
$\Delta$ describing the spin gap at {\bf Q}, $A({\bf
k}\omega)=-\pi^{-1}{\rm Im}\,G({\bf k}\omega)$ and $B({\bf
k}\omega)=-\pi^{-1}{\rm Im}\,D({\bf k}\omega)$ are the hole and spin
spectral functions, $N$ is the number of sites,
$n_F(\omega)=[\exp(\omega/T)+1]^{-1}$,
$n_B(\omega)=[\exp(\omega/T)-1]^{-1}$ with the temperature $T$. Real
parts of the self-energies are calculated from their imaginary parts
and the Kramers-Kronig relation. The self-energy equations for the
superconducting state can be found in Ref.~\cite{Sherman}.

Equations~(\ref{constraint})--(\ref{energies}) form a closed set which
can be solved by iteration for fixed chemical potential $\mu$ and
temperature $T$. We carried out such calculations for the parameters of
the model $J=0.1$~eV, $t=0.5$~eV which correspond to hole-doped
cuprates \cite{McMahan}. To stabilize the iteration procedure an
artificial damping $i\eta$, $\eta=(0.015-0.05)t$, was added to the
frequency in the hole Green's function.

The results of such calculations can be directly compared with the
imaginary part of the magnetic susceptibility $\chi''$ derived from the
neutron scattering experiments, since this quantity is connected with
$B({\bf k}\omega)$,
\begin{equation}
\chi''({\bf k}\omega)=4\pi\mu_B^2 B({\bf k}\omega)
=-\frac{16\mu_B^2(1-\gamma_{\bf k})(J|C_1|+tF_1)\omega{\rm
Im}\,\Pi}{(\omega^2-\omega{\rm Re}\,\Pi-\omega_{\bf k}^2)^2+(\omega{\rm
Im}\,\Pi)^2}, \label{chi}
\end{equation}
where $\mu_B$ is the Bohr magneton. Equation~(\ref{chi}) is the most
general form for the magnetic susceptibility which follows from the
Mori formalism. With the spin excitation damping $\Pi({\bf k}\omega)$
described by the fermion bubble (\ref{se}) this equation acquires some
similarity with the susceptibility derived in the random phase
approximation \cite{Liu} which is frequently used for the discussion of
the magnetic properties of cuprates. For this similarity the term
$\omega^2_{\bf k}$ in the denominator of Eq.~(\ref{chi}) has to be
negligibly small in comparison with $\omega{\rm Re}\Pi({\bf k}\omega)$.
For the hole spectrum self-consistently calculated here this can happen
only in the overdoped region $x\gtrsim 0.3$ when the spin correlation
$C_1$ which determines the magnitude of $\omega_{\bf k}$ [see
Eq.~(\ref{energies})] tends to zero \cite{Sherman,Bonca}.

\section{The spectrum of spin excitations}
As follows from Eq.~(\ref{chi}), the frequencies of spin excitations
satisfy the equation
\begin{equation}\label{magnons}
\omega^2-\omega{\rm Re}\,\Pi({\bf k}\omega)-\omega^2_{\bf k}=0.
\end{equation}
As seen from Eq.~(\ref{se}), ${\rm Im}\,\Pi({\bf k}\omega)$ is finite
for ${\bf k}\rightarrow 0$, whereas $\omega_{\bf k}$ vanishes in this
limit. Therefore the spin Green's function has a purely imaginary,
diffusive pole near the $\Gamma$ point, in compliance with the result
of the hydrodynamic theory \cite{Forster}.

\begin{figure}[ht]
\begin{center}
\includegraphics[width=7cm,keepaspectratio,clip]{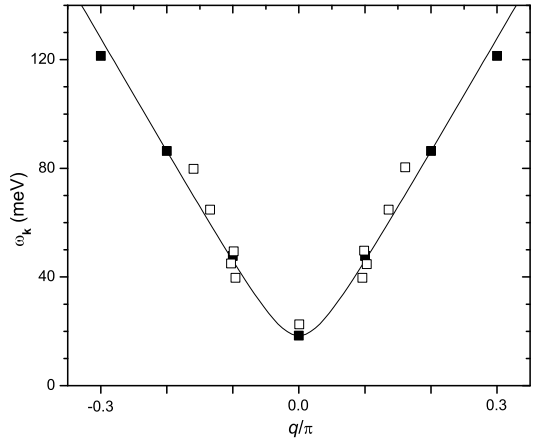}
\end{center}
\caption{\label{Fig_i} The dispersion of spin excitations near {\bf Q}
along the edge of the Brillouin zone. The dispersion was calculated in
a 20$\times$20 lattice for $x=0.06$ and $T=17$~K (filled squares, the
solid line is the fit with $\omega_{\bf k}=[\omega^2_{\bf Q}+c^2({\bf
k-Q})^2]^{1/2}$). Open squares are the experimental dispersion
\protect\cite{Bourges} of the maximum in the frequency dependence of
the odd $\chi''({\bf q}\omega)$, ${\bf q=k-Q}$, in
YBa$_2$Cu$_3$O$_{6.5}$ ($x\approx 0.075$ \cite{Tallon}) at $T=5$~K.}
\end{figure}

The dispersion of spin excitations near {\bf Q} is of special interest,
because this region gives the main contribution into the neutron
scattering. This dispersion is shown in Fig.~\ref{Fig_i}.
As seen from this figure, in contrast to the usual spin-wave theory the
frequency of spin excitations at the antiferromagnetic wave vector {\bf
Q} is finite. This frequency $\omega_{\bf
Q}=4J(2\alpha|C_1|\Delta)^{1/2}$ is directly connected with the spin
correlation length $\xi$. Indeed, using Eq.~(\ref{se}) and taking into
account that the region near {\bf Q} gives the main contribution to the
summation over {\bf k}, we find for large distances and low
temperatures
\begin{equation}\label{spincor}
\langle s^z_{\bf l}s^z_{\bf 0}\rangle=N^{-1}\sum_{\bf k}{\rm e}^{i\bf
kl}\int_0^\infty\!\! d\omega\,\coth\left(\frac{\omega}{2T}\right)
B({\bf k}\omega)\propto{\rm e}^{i\bf Ql}(\xi/|{\bf l}|)^{1/2}{\rm
e}^{-|{\bf l}|/\xi},
\end{equation}
where $\xi=\frac{1}{2}a\Delta^{-1/2}$ and $a$ is the distance between
in-plane Cu sites. As follows from the calculations, for low
concentrations and temperatures $\Delta\approx 0.2x$ and consequently
$\xi\approx ax^{-1/2}$. An analogous relation between the spin
correlation length and the concentration has been derived from
experimental data in La$_{2-x}$Sr$_x$CuO$_4$ \cite{Keimer}.

In accord with the Hohenberg-Mermin-Wagner theorem \cite{Mermin} the
considered two-dimensional system is in the paramagnetic state for
$T>0$. This result can also be obtained using Eq.~(\ref{se}). Also it
can be shown \cite{Sherman} that in the infinite crystal at $T=0$ the
long-range antiferromagnetic order is destroyed for the hole
concentrations $x>x_c\approx 0.02$.

\section{The resonance peak}
\begin{figure}[ht]
\includegraphics[width=.47\textwidth]{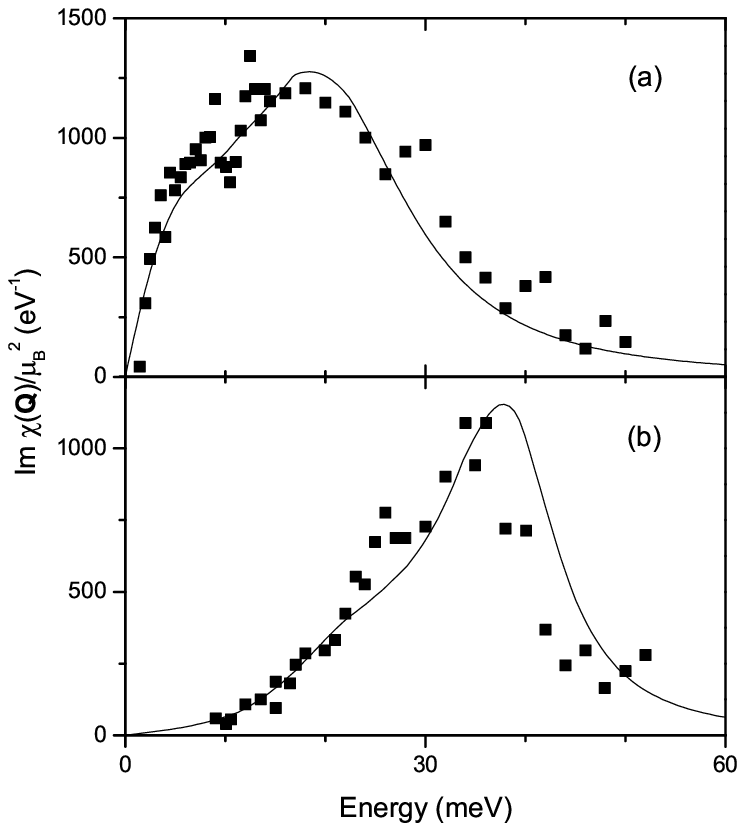}\hspace{1em}
\includegraphics[width=.47\textwidth]{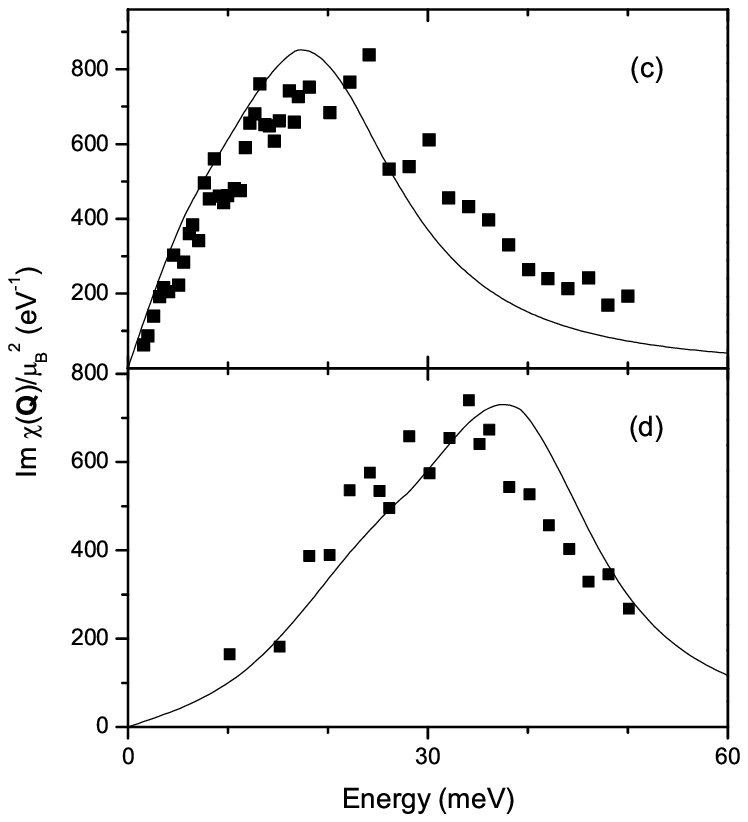}
\caption{The imaginary part of the spin susceptibility at {\bf Q} in
the $d$-wave superconducting (a,b) and normal (c,d) states. Curves show
the results of our calculations for $\eta=0.015t$, $T=17$~K (a,b) and
$T=116$~K (c,d), $x=0.06$ (a,c) and $x=0.12$ (b,d). Squares are the odd
susceptibility measured \protect\cite{Bourges} in
YBa$_2$Cu$_3$O$_{6.5}$ ($T_c=45$~K, $x\approx 0.075$, a and c) and in
YBa$_2$Cu$_3$O$_{6.83}$ ($T_c=85$~K, $x\approx 0.14$
\protect\cite{Tallon}, b and d) at $T=5$~K (a,b) and at $T=100$~K
(c,d). To depict the experimental data in the same scale they were
multiplied by the factor $1.5$.}\label{Fig_ii}
\end{figure}

The imaginary part of the spin susceptibility at the antiferromagnetic
wave vector, obtained in our calculations, is shown in
Fig.~\ref{Fig_ii}.
The experimental data on $\chi''$ in YBa$_2$Cu$_3$O$_{7-y}$
\cite{Bourges} are also depicted here. YBa$_2$Cu$_3$O$_{7-y}$ is a
bilayer crystal and the symmetry allows one to divide the
susceptibility into odd and even parts. The odd part, which for the
antiferromagnetic intrabilayer coupling can be compared with our
single-layer calculations, is shown in the figure.

The maximum in Fig.~\ref{Fig_ii} is the resonance peak. As seen from
this figure, the model reproduces satisfactorily the evolution of the
peak position and shape with doping and temperature. In the $t$-$J$
model the peak is connected with the coherent excitations of localized
Cu spins near the antiferromagnetic wave vector {\bf Q} and the peak
frequency for ${\bf k=Q}$ is close to the frequency $\omega_{\bf Q}$ in
Fig.~\ref{Fig_i}. It is also seen in this figure that the experimental
dispersion of the peak is close to the dispersion of the mentioned spin
excitations. Thus, we came to conclusion that the resonance peak in
YBa$_2$Cu$_3$O$_{7-y}$ and apparently in other cuprate perovskites
where it was observed is a manifestation of the magnon branch modified
in the short-range antiferromagnetic order.

The above results are related to the underdoped case where the
resonance peak is observed both in the normal and superconducting
states. In the overdoped case in YBa$_2$Cu$_3$O$_{7-y}$ the peak is
observed in the superconducting state only \cite{Bourges}. Calculations
for the $t$-$J$ model show that in the normal state on approaching the
overdoped region the maximum in $\chi''$ rapidly loses its intensity,
broadens and shifts to higher frequencies \cite{Sherman04a}. This
result correlates with the broad feature observed experimentally in
these conditions. The mechanism of the peak reappearance in the
superconducting state in the overdoped region was considered in
Ref.~\cite{Morr}. The opening of the $d$-wave superconducting gap with
the magnitude $2\Delta^s>\omega_{\bf Q}$ decreases considerably the
spin excitation damping near {\bf Q} which restores the peak in
$\chi''$.

As seen from Fig.~\ref{Fig_ii}, the resonance peak has a low-frequency
shoulder which is more pronounced for low concentrations and
temperatures. The shoulder stems from the frequency dependence of the
magnon damping ${\rm Im}\,\Pi({\bf Q}\omega)$ which in its turn is a
consequence of the hole Fermi surface nesting existing in the $t$-$J$
model for moderate doping \cite{Sherman}. An analogous nesting is
expected in the two-layer YBa$_2$Cu$_3$O$_{7-y}$ between the parts of
the Fermi surface belonging to the bonding and antibonding bends
\cite{Liu}.

\begin{figure}
\includegraphics[width=.47\textwidth]{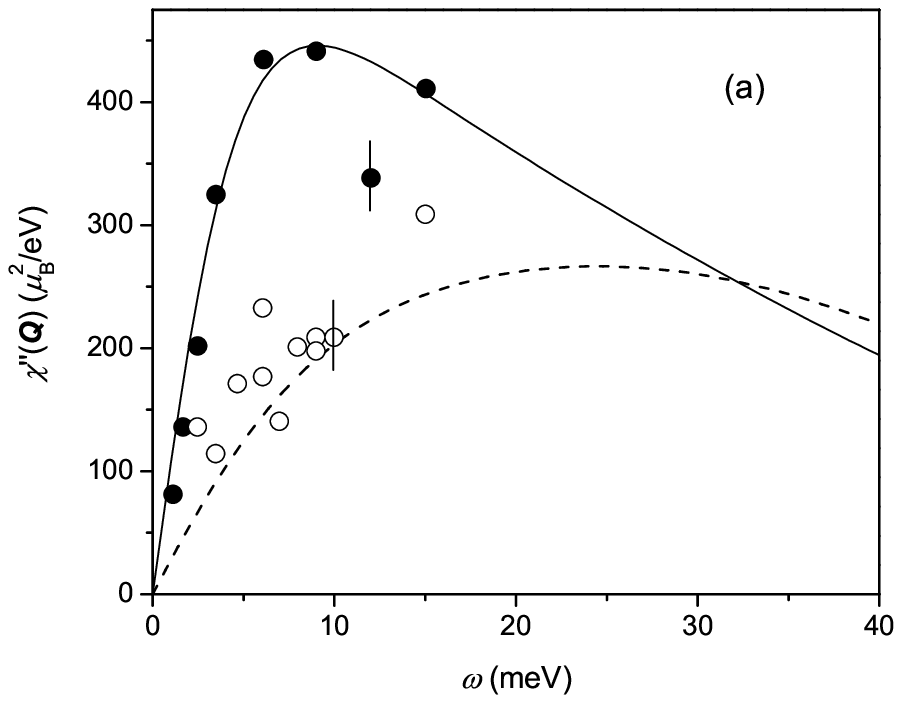}\hspace{1em}
\includegraphics[width=.47\textwidth]{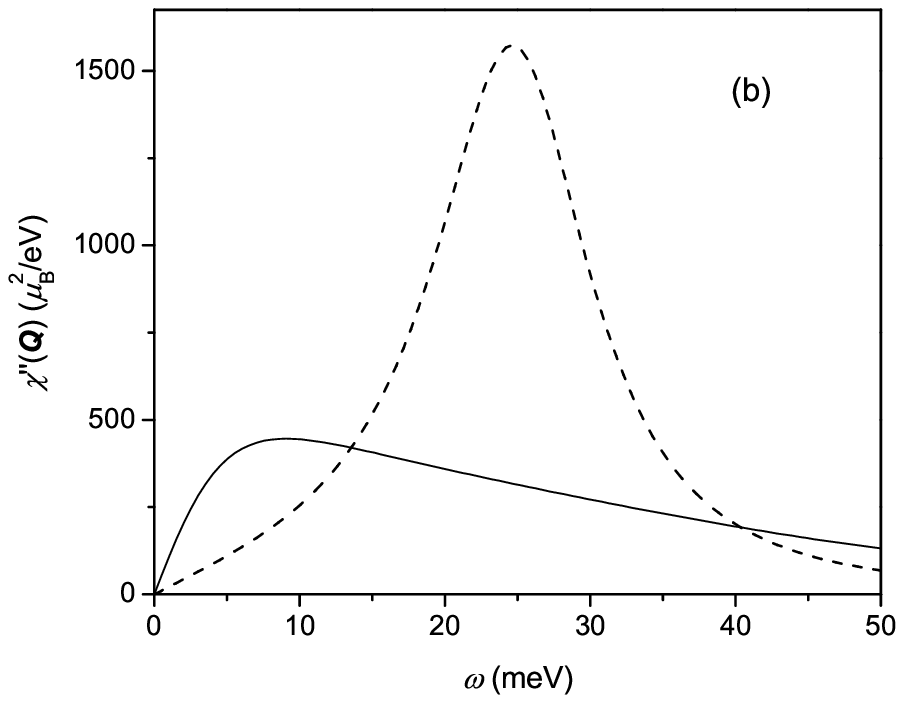} \caption{The frequency
dependence of the imaginary part of the susceptibility. (a) The solid
line corresponds to $T=29$~K, $\eta=0.05t$ and $x \approx 0.1$. The
dashed line is for $T=116$~K and the same other parameters. Symbols are
experimental results \protect\cite{Aeppli} in
La$_{1.86}$Sr$_{0.14}$CuO$_4$ at $T=35$~K (filled circles) and $T=80$~K
(open circles). Vertical bars show experimental errors. The calculated
susceptibility is given for ${\bf k}={\bf Q}$, the experimental data
are for a wave vector of the incommensurate peak. (b) The solid line is
the same as in part (a). The dashed line is calculated for
$\eta=0.015t$, other parameters are the same as for the solid line.}
\label{Fig_iii}
\end{figure}

It is worth noting that for all four curves in Fig.~\ref{Fig_ii} the
value of $|\Pi({\bf Q},\omega_{\bf Q})|/2$ is smaller than $\omega_{\bf
Q}$. The satisfactory agreement between the experimental and calculated
results allows us to conclude that near {\bf Q} the spin excitations
are not overdamped in underdoped YBa$_2$Cu$_3$O$_{7-y}$. In calculating
the curves in Fig.~\ref{Fig_ii} the artificial broadening $\eta$ in the
hole Green's function was set equal to $0.015t$. This broadening
simulates contributions to the hole damping from mechanisms differing
from the spin excitation scattering. The magnitude of these
interactions can essentially vary in different crystals. The example of
the magnetic susceptibility calculated with the larger hole damping
$\eta=0.05t$ is shown in Fig.~\ref{Fig_iii}.
As seen from this figure, the spin excitation damping is very sensitive
to the hole damping. The increase of this latter damping leads to a
marked growth of the spin excitation damping which results in the
overdamping of these excitations. Instead of the pronounced peak at the
frequency $\omega_{\bf Q}$ a broad low-frequency feature is observed in
the susceptibility (see Fig.~\ref{Fig_iii}b). This shape resembles that
observed in La$_{2-x}$Sr$_x$CuO$_4$ \cite{Aeppli}. The comparison was
carried out in Fig.~\ref{Fig_iii}a. It should be noted that the use of
a comparatively small 20$\times$20 lattice did not allow us to describe
the incommensurability of the magnetic response -- $\chi''$ is peaked
at {\bf Q} and our calculated data belong to this momentum (the use of
a larger lattice is too time-consuming for the self-consistent
calculations). In La$_{1.86}$Sr$_{0.14}$CuO$_4$ the low-frequency
susceptibility is peaked at incommensurate momenta ${\bf k}=(\pi\pm
2\pi\delta,\pi), (\pi,\pi\pm 2\pi\delta)$ and the experimental data
\cite{Aeppli} shown in Fig.~\ref{Fig_iii}a belong to one of these
momenta. These data were increased by approximately 2.5 times to show
them in the scale with the calculated results. The need for scaling is
apparently connected with the splitting of the commensurate peak into
the four incommensurate maxima. Apart from this difference in the
momentum dependencies, the calculated frequency and temperature
dependencies are in good agreement with the experimental results. Thus,
it can be concluded that the dissimilarity of the frequency
dependencies of the susceptibility in YBa$_2$Cu$_3$O$_{7-y}$ and
La$_{2-x}$Sr$_x$CuO$_4$ is connected with different values of the
damping of spin excitations, which are well defined at the
antiferromagnetic wave vector in the former crystal and overdamped in
the latter. This property of spin excitations in
La$_{2-x}$Sr$_x$CuO$_4$ is not changed in the superconducting state due
to the small superconducting gap $2\Delta^s\approx 9{\rm meV}<
\omega_{\bf Q}$ in this crystal \cite{Morr}.

\section{Incommensurability in the magnetic response}
\begin{figure}
\centerline{\includegraphics[width=4cm]{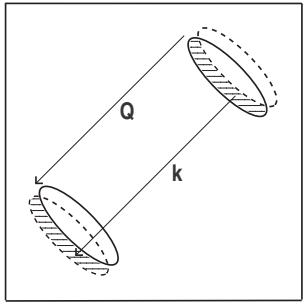}} \caption{The
Brillouin zone of the square lattice. Solid curves are two of four
ellipses forming the Fermi surface at small $x$. Dashed lines are the
Fermi surface contours shifted by $\pm({\bf k-Q})$. Regions of ${\bf
k'}$ and ${\bf k+k'}$ contributing to the damping
(\protect\ref{apprdamping}) are shaded.}\label{Fig_iv}
\end{figure}

For low frequencies susceptibility (\ref{chi}) is essentially
simplified,
\begin{equation}\label{chiloww}
\chi''({\bf k}\omega)\propto\frac{\omega{\rm Im}\Pi({\bf
k}\omega)}{\omega^4_{\bf k}}.
\end{equation}
As seen from Fig.~\ref{Fig_i} and Eq.~(\ref{energies}), $\omega_{\bf
k}$ is the increasing function of the difference ${\bf k-Q}$ and
therefore the denominator of the fraction (\ref{chiloww}) favors the
appearance of a commensurate peak centered at the antiferromagnetic
momentum in the susceptibility. However, if the spin excitation damping
has a pronounced dip at ${\bf k=Q}$ the peak splits into several peaks
shifted from {\bf Q}. To make sure that the damping may really have
such a behavior let us consider the case of low hole concentrations and
temperatures. In this case the hole Fermi surface consists of four
ellipses centered at $(\pm\pi/2,\pm\pi/2)$ \cite{Sherman}. Two of them
are shown in Fig.~\ref{Fig_iv}.
The spin excitation damping described by the fermion bubble (\ref{se})
is simplified to
\begin{eqnarray}
{\rm Im}\,\Pi({\bf k}\omega)&\propto&\sum_{\bf k'}(\gamma_{\bf
k'}-\gamma_{\bf k'+k})^2 \Big[n_F(\varepsilon_{\bf
k'+k}-\mu)-n_F(\varepsilon_{\bf k'}-\mu)\Big]\nonumber\\
&\times&\delta(\omega+\varepsilon_{\bf k'}-\varepsilon_{\bf
k'+k})\label{apprdamping}
\end{eqnarray}
in the considered case. Here $\varepsilon_{\bf k}$ is the hole
dispersion. In the process described by Eq.~(\ref{apprdamping}) a
fermion picks up an energy and momentum of a defunct spin excitation
and is transferred from a region below the Fermi level to a region
above it. However, for ${\bf k=Q}$ this process is impossible, because
for initial states ${\bf k'}$ interior to an ellipse in
Fig.~\ref{Fig_iv} all final states with momenta ${\bf k+k'}$ will be
inside another ellipse, i.e. below the Fermi level. Thus, in this
simplified picture the damping vanishes for ${\bf k=Q}$ and grows with
increasing the difference ${\bf k-Q}$ approximately proportional to the
shaded area in Fig.~\ref{Fig_iv} \cite{Sherman04b}.

With increasing the hole concentration the Fermi surface of the $t$-$J$
model is transformed to a rhombus centered at {\bf Q}
\cite{Sherman04a}. This result is in agreement with the Fermi surface
observed in La$_{2-x}$Sr$_x$CuO$_4$ \cite{Zhou} [however, to reproduce
the experimental Fermi surface terms describing the hole transfer to
more distant coordination shells have to be taken into account in the
kinetic term of Hamiltonian (\ref{hamiltonian})]. For such $x$ another
mechanism of the dip formation in the damping comes into effect. The
interaction constant $\gamma_{\bf k'}-\gamma_{\bf k'+k}$ in
Eq.~(\ref{apprdamping}) vanishes in the so-called ``hot spots'' -- the
crossing points of the Fermi surface with the boundary of the magnetic
Brillouin zone. For small frequencies and ${\bf k=Q}$ the nearest
vicinity of these points contributes to the damping. Consequently, the
damping has a dip at {\bf Q}. With the inclusion of the hole transfer
to more distant coordination shells the interaction constant is
changed, however the conclusion about the dip at the antiferromagnetic
momentum in the damping remains unchanged.

The dip disappears when the hole damping $\eta$ exceeds the frequency
$\omega>0$. Since the main contribution to the damping is made by
states with energies $-\omega\leq\varepsilon_{\bf k'}-\mu\leq 0$ and
$0\leq\varepsilon_{\bf k'+k}-\mu\leq\omega$, this means that the above
consideration is valid when there exist well defined quasiparticle
excitations near the Fermi surface.

\begin{figure}[htb]
\includegraphics[width=.47\textwidth]{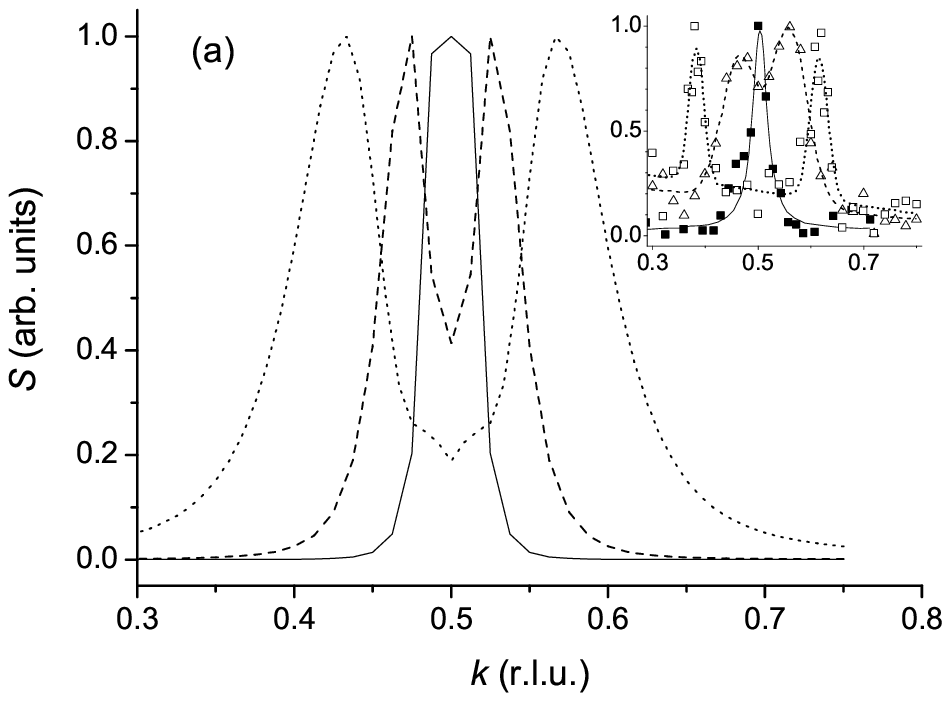}\hspace*{1em}
\raisebox{1ex}{\includegraphics[width=.47\textwidth]{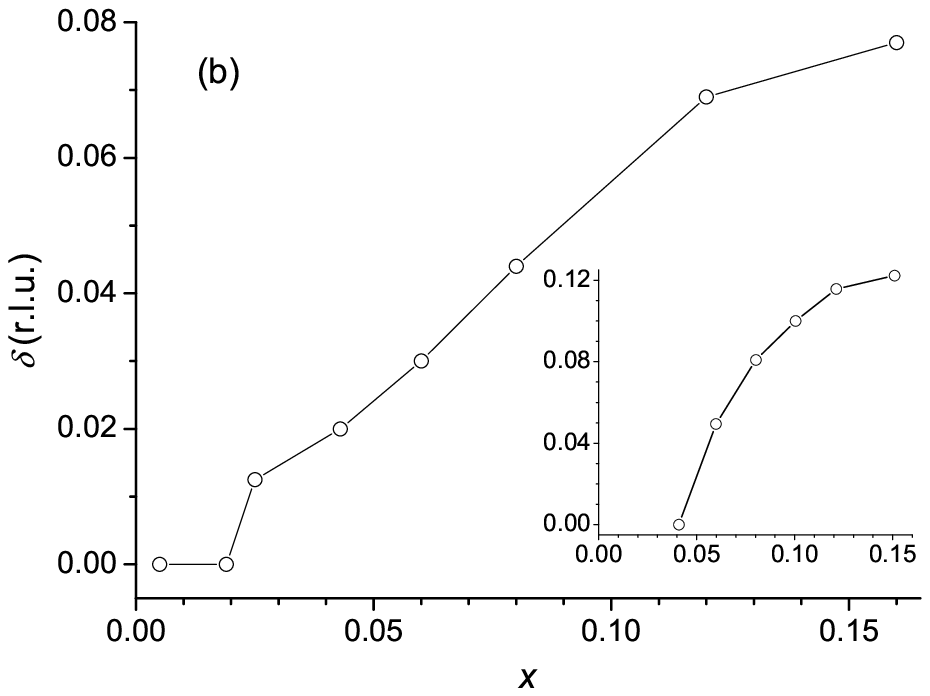}}
\caption{(a) The dynamic structure factor vs.\ wave vector along the
edge of the Brillouin zone for $x=0.015$ (solid line), 0.06 (dashed
line), and 0.12 (short dash). For all curves $T=58$~K and
$\omega=2$~meV. Inset: the structure factor measured
\protect\cite{Yamada,Wakimoto} in La$_{2-x}$Sr$_x$CuO$_4$ for $x=0.03$,
$T=2$~K, $\omega=0$ (filled squares and solid line), $x=0.06$,
$T=25$~K, $\omega=2$~meV (triangles and dashed line), and $x=0.12$,
$T=31$~K, $\omega=2$~meV (open squares and short dash). Fitting curves
in the inset are from Refs.~\protect\cite{Yamada,Wakimoto}. (b) The
incommensurability parameter $\delta$ vs.\ $x$ for $T=58$~K and
$\omega=2$~meV. Inset: experimental data \protect\cite{Yamada} for
La$_{2-x}$Sr$_x$CuO$_4$. The wave vector and the incommensurability
parameter are given in the reciprocal lattice units
$2\pi/a$.}\label{fig_v}
\end{figure}

Results of our calculations in a 120$\times$120 lattice and
experimental data are shown in Fig.~\ref{fig_v}.
In agreement with experiment \cite{Yamada,Yoshizawa} in our
calculations peaks in the dynamic structure factor $S({\bf
k}\omega)=[1-\exp(-\omega/T)]^{-1}\chi''({\bf k}\omega)$ along the edge
of the Brillouin zone [the direction $(\pi,\pi)-(0,\pi)$] are more
intensive than those along the diagonal [the direction
$(\pi,\pi)-(0,0)$]. As seen from Fig.~\ref{fig_v}, the calculations
reproduce correctly the main known features of the incommensurate
response except that the experimental $\delta$ is somewhat larger than
the calculated one. In experiment with increasing $T$ the
incommensurability decreases and finally disappears. The calculations
reproduce this peculiarity also. In the calculations the concentration
dependence of $\delta$ is mainly determined by the spin gap parameter
$\Delta$ in $\omega_{\bf k}$ in Eq.~(\ref{energies}). For $x<0.12$ it
grows with $x$ and saturates at larger concentrations \cite{Sherman}.

With increasing $\omega$ the dip in the spin excitation damping is
shallowed and finally disappears. Besides, on approaching $\omega_{\bf
Q}$ the denominator in Eq.~(\ref{chi}) will favor the appearance of the
commensurate peak at {\bf Q}. Thus, the low-frequency incommensurate
maxima of $\chi''$ converge to the commensurate peak at
$\omega\approx\omega_{\bf Q}$. The dispersion of the maxima in $\chi''$
above $\omega_{\bf Q}$ is determined by the denominator in
Eq.~(\ref{chi}) and is close to that shown in Fig.~\ref{Fig_i}.
Consequently, the dispersion of the maxima in $\chi''$ resembles two
parabolas converging near the point $({\bf Q},\omega_{\bf Q})$. The
upper parabola with branches pointed up reflects the dispersion of spin
excitations, while the lower parabola with branches pointed down stems
from the momentum dependence of the spin excitation damping. Such kind
of the dispersion is indeed observed in cuprates \cite{Arai,Tranquada}.

Generally the incommensurate magnetic response is not accompanied by an
inhomogeneity of the carrier density. In works based on the stripe
mechanism \cite{Tranquada} the magnetic incommensurability is connected
with the appearance of a charge density wave. Notice however, that the
magnetic incommensurability is observed in all lanthanum cuprates,
while the charge density wave is detected in neutron scattering in
those cuprates which are in the low-temperature tetragonal or
less-orthorhombic phases (La$_{2-x}$Ba$_x$CuO$_4$,
La$_{2-y-x}$Nd$_y$Sr$_x$CuO$_4$) \cite{Fujita}. In these phases the
charge density wave is stabilized by the corrugated pattern of the
in-plane lattice potential. It can be supposed that the magnetic
incommensurability may be a precursor rather than a consequence of the
charge density wave.

\section{Concluding remarks}
In this paper the projection operator technique was used for
investigating the magnetic properties of the $t$-$J$ model of cuprate
high-$T_c$ superconductors. It was demonstrated that the calculations
reproduce correctly the frequency and momentum dependencies of the
experimental magnetic susceptibility and its variation with doping and
temperature in the normal and superconducting states in
YBa$_2$Cu$_3$O$_{7-y}$ and lanthanum cuprates. This comparison with
experiment allowed us to associate the resonance peak in
YBa$_2$Cu$_3$O$_{7-y}$ with the magnon branch modified in the
short-range antiferromagnetic order. The lack of the resonance peak in
La$_{2-x}$Sr$_x$CuO$_4$ was connected with an increased damping of spin
excitations at the antiferromagnetic wave vector ${\bf Q}=(\pi,\pi)$.
One of the possible reasons for the overdamped excitations is an
increased hole damping in this crystal. It was shown that for low
frequencies the susceptibility is peaked at incommensurate momenta
$(\pi\pm 2\pi\delta,\pi)$ and $(\pi,\pi\pm 2\pi\delta)$. The
incommensurability is the consequence of the dip in the momentum
dependence of the spin excitation damping at {\bf Q}. The dip appears
due to the Fermi surface nesting for low hole concentrations $x$ and
due to small hole-magnon interaction constants for moderate
concentrations. In agreement with experiment for $x \lesssim 0.12$ the
incommensurability grows nearly proportional to $x$ and tends to
saturation for $x>0.12$. This is connected with the concentration
dependence of the spin excitation frequency at {\bf Q}. Also in
agreement with experiment the incommensurability decreases with
increasing temperature. The dispersion of the maxima in the
susceptibility resembles two converging parabolas. The upper parabola
with branches pointed up reflects the dispersion of spin excitations,
while the lower parabola with branches pointed down stems from the
momentum dependence of the spin excitation damping. In the considered
mechanism the magnetic incommensurability is not accompanied by the
inhomogeneity of the carrier density.

\begin{acknowledgments}
This work was supported by the ESF grant No.~5548.
\end{acknowledgments}

\end{document}